\documentclass{article}
\usepackage{spconf,amsmath,graphicx}
\usepackage{booktabs}
\usepackage{multirow}
\usepackage{bbding}
\usepackage{hyperref}
\usepackage{amssymb}
\usepackage{url}

\usepackage{fancyhdr}

\title{A two-stage framework in cross-spectrum domain for real-time speech enhancement}
\name{Yuewei Zhang$^{1}$ \qquad Huanbin Zou$^{2}$ \qquad Jie Zhu$^{1}$\thanks{The first two authors equally contribute to this work. Jie Zhu is the corresponding author (Email: zhujie@sjtu.edu.cn).}}
\address{$^{1}$ Department of Electronic Engineering, Shanghai Jiao Tong University, Shanghai, China \\
${^2}$ Tencent Video Cloud, Shanghai, China}
\begin{document}
%
\maketitle

\thispagestyle{fancy}
\fancyhead{}
\lhead{}
\lfoot{\small \centering © 20XX IEEE. Personal use of this material is permitted. Permission from IEEE must be obtained for all other uses, in any current or future media, including reprinting/republishing this material for advertising or promotional purposes, creating new collective works, for resale or redistribution to servers or lists, or reuse of any copyrighted component of this work in other works.}
\cfoot{}
\rfoot{}

\begin{abstract}
Two-stage pipeline is popular in speech enhancement tasks due to its superiority over traditional single-stage methods. The current two-stage approaches usually enhance the magnitude spectrum in the first stage, and further modify the complex spectrum to suppress the residual noise and recover the speech phase in the second stage. The above whole process is performed in the short-time Fourier transform (STFT) spectrum domain. In this paper, we re-implement the above second sub-process in the short-time discrete cosine transform (STDCT) spectrum domain. The reason is that we have found STDCT performs greater noise suppression capability than STFT. Additionally, the implicit phase of STDCT ensures simpler and more efficient phase recovery, which is challenging and computationally expensive in the STFT-based methods. Therefore, we propose a novel two-stage framework called the ST\textbf{F}T-ST\textbf{D}CT spectrum \textbf{f}usion \textbf{net}work (\textbf{FDFNet}) for speech enhancement in cross-spectrum domain. Experimental results demonstrate that the proposed FDFNet outperforms the previous two-stage methods and also exhibits superior performance compared to other advanced systems.
\end{abstract}
\section{Introduction}
\label{sec:intro}


Recently, with the rapid development of deep learning (DL), a lot of DL-based speech enhancement (SE) methods \cite{defossez20_interspeech,8547084,9054597,hu20g_interspeech,9431717} are proposed. It has been illustrated that the DL-based methods have better performance than the traditional ones. The mainstream SE methods process the speech signal in the time-frequency (TF) domain \cite{8547084,9054597,hu20g_interspeech,9431717}. Specifically, the noisy speech waveform is firstly converted to a TF spectrum by TF transformation. Then, it is fed into a deep neural network (DNN), which is trained to predicted the target speech’s spectrum or the corresponding spectrum mask. Finally, the enhanced speech is reconstructed by the TF inverse transformation. In the existing works, short-time Fourier transform (STFT) is the most common TF transformation method.

For a long time, the TF domain methods only focus on recovering the target magnitude spectrum but leaving the phase information unchanged \cite{tan18_interspeech}. However, it is latter demonstrated that an accurate phase recovery can further improve the SE performance \cite{PALIWAL2011465}. As a result, the phase-aware SE methods \cite{hu20g_interspeech,Yin_Luo_Xiong_Zeng_2020} have thrived in the past few years. Since the speech’s phase spectrum does not exhibit obvious structural features like the magnitude spectrum, many methods \cite{hu20g_interspeech,choi2018phaseaware} attempt to indirectly predict the phase in the complex domain. In another word, they simultaneously enhance the real and imaginary parts of noisy spectrum. Nevertheless, it is still challenging to construct a DNN to accurately predict the target complex spectrum in one stage. To alleviate this problem, two-stage algorithm \cite{9431717,li21g_interspeech} is proposed to decompose the original single-stage optimization task into two easier and progressive sub-tasks. Specifically, the first stage is responsible for magnitude spectrum enhancement, so as to coarsely remove the noise. Subsequently, the second stage further predicts the residual component of the complex spectrum, so as to suppress the residual noise and recover the speech phase. The experiments have demonstrated that two-stage algorithm outperforms the traditional single-stage approach.

\begin{figure*}[htbp]
\centerline{\includegraphics[height=1.28in]{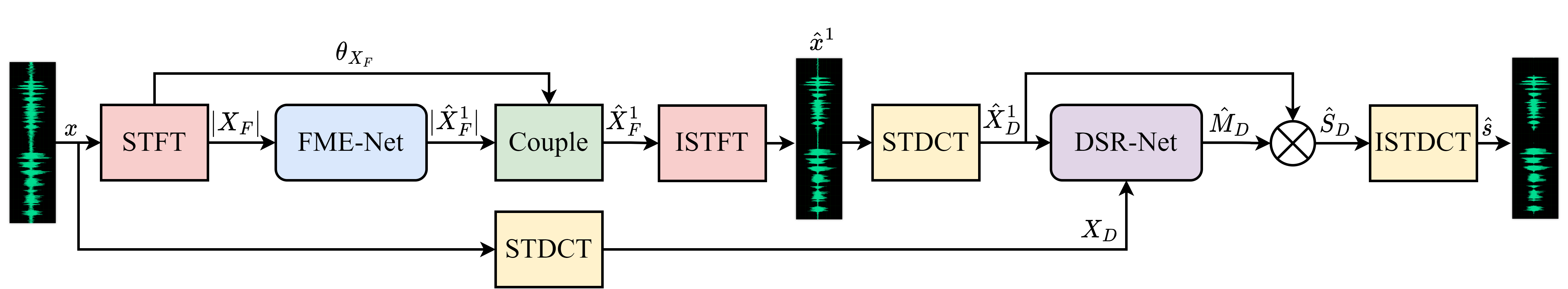}}
\caption{Overall structure of the ST\textbf{F}T-ST\textbf{D}CT spectrum \textbf{f}usion \textbf{net}work (FDFNet).} \vspace{-0.35cm} \label{fig1}
\end{figure*}

However, the previous two-stage methods have some shortcomings:
(1) The explicit phase estimation in the STFT spectrum domain is still challenging. (2) After enhancing the STFT magnitude spectrum, it is difficult to further suppress the residual noise in the STFT complex spectrum domain.

In this paper, we propose to improve the existing two-stage algorithms. We employ the short-time discrete cosine transform (STDCT) \cite{1672377} instead of the previously used STFT to construct the second stage’s model. Since STDCT is a real-valued TF transformation, both the magnitude and phase information coexist in a real-valued spectrum. Therefore, the STDCT-based model in our improved two-stage algorithm can implicitly recover the clean phase. Meanwhile, we find that the STDCT-based SE model has a stronger noise reduction capability than the STFT-based SE model, which enables better residual noise suppression in the second stage. To the best of our knowledge, this is the first study that combines both STFT and STDCT to realize SE in the cross-spectrum domain. We name our proposed method the ST\textbf{F}T-ST\textbf{D}CT spectrum \textbf{f}usion network (\textbf{FDFNet}), and it adopts a causal configuration to ensure real-time SE. The experimental results demonstrate the effectiveness and superiority of our method.

\section{Proposed Method}
\label{sec:method}
\subsection{The overall architecture}
Our FDFNet adopts the two-stage framework, and it generally consists of a ST\textbf{F}T-based \textbf{m}agnitude \textbf{e}nhancement sub-\textbf{net}work (dubbed FME-Net) and a ST\textbf{D}CT-based \textbf{s}pectrum \textbf{r}efinement sub-\textbf{net}work (dubbed DSR-Net). The overall architecture of FDFNet is illustrated in Fig.\ \ref{fig1}. The model input is the noisy speech $x$, which can be expressed as:
\begin{equation}
x=s+n
\label{equation1}
\end{equation}
where $s$ and $n$ represent the clean speech and noise, respectively. 

In the first stage, we adopt the typical magnitude-only convolutional recurrent network (CRN) \cite{tan18_interspeech} as FME-Net, which includes the convolutional encoder, recurrent neural network (RNN), and deconvolutional decoder. In detail, the convolutional encoder consists of several 2D convolutional (Conv2d) blocks, each of which is composed of a 2D convolutional layer followed by batch normalization and PReLU. For the deconvolutional decoder, it is the symmetric version of the encoder, where each 2D convolutional layer is replaced by the 2D deconvolutional layer. Between the encoder and decoder, several gated recurrent unit (GRU) layers are inserted to model the temporal correlations. In addition, skip connection is utilized to connect each encoder block to its corresponding decoder block for better performance. FME-Net receives the noisy speech’s magnitude spectrum $|X_F|$, which is derived by STFT, and estimates the enhanced magnitude spectrum $|\hat{X}^1_F|$. Then, we couple this enhanced magnitude $|\hat{X}^1_F|$ with the original noisy phase $\theta_{X_F}$ to obtain the enhanced STFT spectrum $\hat{X}^1_F$ of the first stage. This stage aims at suppressing the noise components coarsely.

After the FME-Net, we sequentially apply inverse STFT (ISTFT) and STDCT to $\hat{X}^1_F$, so as to obtain the pre-enhanced STDCT spectrum $\hat{X}^1_D$. Since STDCT is a real-valued TF transformation, the spectrum $\hat{X}^1_D$ contains both the magnitude and phase information. Therefore, the following DSR-Net which directly operates on the STDCT spectrum can simultaneously recover the target magnitude and phase.

In the second stage, DSR-Net takes both the pre-enhanced STDCT spectrum $\hat{X}^1_D$ and the original noisy STDCT spectrum $X_D$ as the input. Similar to previous works \cite{9431717,li21g_interspeech}, the output of the second stage is only used to make a further refinement to the first stage's result. The details of our DSR-Net will be described in the following section. Overall, DSR-Net predicts a  STDCT spectrum mask $\hat{M}_D$, which is subsequently multiplied with the pre-enhanced STDCT spectrum $\hat{X}^1_D$ to derive the final estimation $\hat{S}_D$ for the target STDCT spectrum.

In a nutshell, the whole procedure of our FDFNet can be formulated as:
\begin{equation}
X_F=|X_F|\cdot{e^{j\theta_{X_F}}}=\mathrm{STFT}(x)
\label{equation2}
\end{equation}
\begin{equation}
|\hat{X}^1_F|=\mathcal{F}_1(|X_F|;\Phi_1)
\label{equation3}
\end{equation}
\begin{equation}
\hat{X}^1_F=|\hat{X}^1_F|e^{j\theta_{X_F}}
\label{equation4}
\end{equation}
\begin{equation}
\hat{X}^1_D=\mathrm{STDCT}(\mathrm{ISTFT}(\hat{X}^1_F))
\label{equation5}
\end{equation}
\begin{equation}
X_D=\mathrm{STDCT}(x)
\label{equation6}
\end{equation}
\begin{equation}
\hat{M}_D=\mathcal{F}_2(X_D,\hat{X}^1_D;\Phi_2)
\label{equation7}
\end{equation}
\begin{equation}
\hat{s}=\mathrm{ISTDCT}(\hat{S}_D)=\mathrm{ISTDCT}(\hat{M}_D\cdot{\hat{X}^1_D})
\label{equation8}
\end{equation}
where $\mathcal{F}_1$ and $\mathcal{F}_2$ represent the functions of FME-Net and DSR-Net with parameter sets $\Phi_1$ and $\Phi_2$, respectively.

\subsection{STDCT-based spectrum refinement sub-network (DSR-Net)}
\begin{figure}[htbp]
\centerline{\includegraphics[height=3.8in]{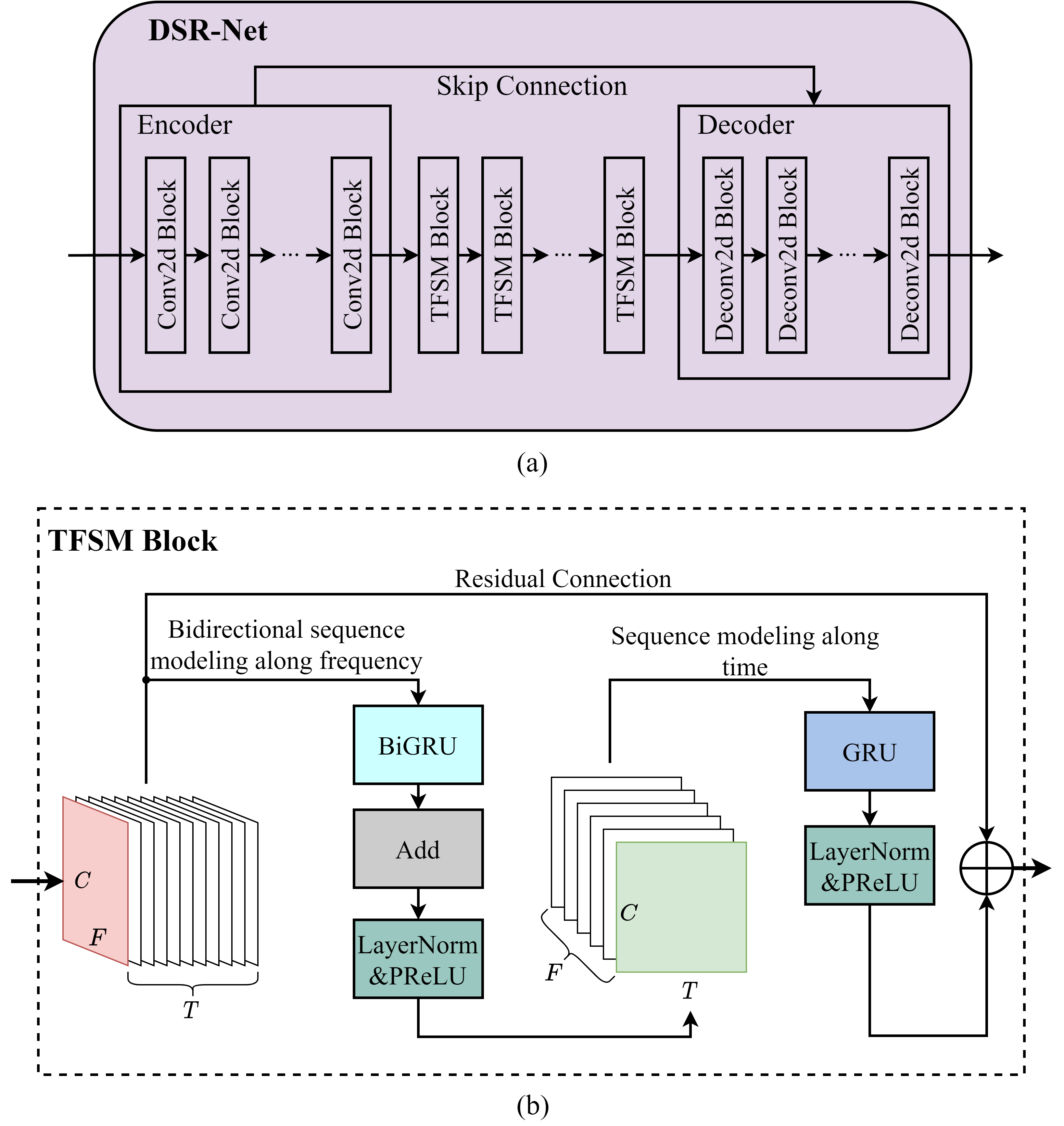}}
\caption{(a) The details of ST\textbf{D}CT-based \textbf{s}pectrum \textbf{r}efinement sub-\textbf{net}work (DSR-Net). (b) The details of time-frequency sequence modeling (TFSM) block.} \vspace{-0.35cm} \label{fig2}
\end{figure}

Considering that the CRN structure has been proven to be effective in the SE tasks \cite{hu20g_interspeech,tan18_interspeech}, we still follow this network topology when designing our DSR-Net. The details of our DSR-Net is illustrated in Fig.\ \ref{fig2}(a). 

Since STDCT spectrum is the real-valued spectrum like the STFT magnitude spectrum, the structure of the encoder and decoder in our DSR-Net is exactly the same as that of FME-Net. While between the encoder and decoder, we design a time-frequency sequence modeling (TFSM) block, as shown in Fig.\ \ref{fig2}(b), to replace the ordinary RNN layer. Similar to the dual-path RNN architecture \cite{9054266}, the TFSM block aims to respectively model the sequential dependencies along the time and frequency dimensions. Specifically, the TFSM block firstly captures the local and global context features among different frequency bins at each frame. This is achieved by a bidirectional GRU (BiGRU) layer, followed by an add layer, layer normalization, and PReLU. The add layer after BiGRU is used to sum up the bidirectional outputs of BiGRU. Then, another GRU layer is applied to the previously processed result to model the temporal correlations. The GRU layer is also followed by layer normalization and PReLU. Residual connection is added between the original input and the sequence modeling result to yield the final output of TFSM block. Multiple TFSM blocks are stacked to ensure the performance. 

The prediction target of our DSR-Net is the DCT ideal ratio mask (DCTIRM), which is defined as:
\begin{equation}
M_D=\frac{S_D}{\hat{X}^1_D}
\label{equation9}
\end{equation}
where $S_D$ is the clean speech's STDCT spectrum. Eventually, the enhanced speech $\hat{s}$ can be obtained according to Eq. (\ref{equation8}).

\subsection{Loss function}
Similar to the previous works \cite{9431717,li21g_interspeech}, We also adopt the two-stage training scheme to optimize our FDFNet. First, we train FME-Net with the mean square error (MSE) loss toward magnitude spectrum estimation, which can be expressed as:
\begin{equation}
\mathcal{L}_{\mathrm{FME}}=\left\||\hat{X}^1_F|-|S_F|\right\|_F^2
\label{equation10}
\end{equation}
where $|S_F|$ is the magnitude spectrum of the clean speech $s$.

Subsequently, we freeze the optimized FME-Net and train the DSR-Net by a hybrid loss:
\begin{equation}
\begin{aligned}
    \mathcal{L}_{\mathrm{DSR}} &= \mathcal{L}_{\mathrm{T}}(\hat{s},s)+\mathcal{L}_{\mathrm{TF}}(\hat{M}_D,M_D) \\
    & = \left\|\hat{s}-s\right\|_1+\left\|\hat{M}_D-M_D\right\|_F^2
\end{aligned}
\label{equation11}
\end{equation}
where $\mathcal{L}_{\mathrm{T}}(\hat{s},s)$ represents the L1 loss between the enhanced speech and clean speech, and $\mathcal{L}_{\mathrm{TF}}(\hat{M}_D,M_D)$ denotes the MSE loss toward DCTIRM estimation.

\section{Experiments and Results}
\label{sec:exp}
\subsection{Experimental setting}

To evaluate the performance of our FDFNet, we conduct the experiments on the VoiceBank+DEMAND dataset \cite{valentini2016investigating}. It includes 11,572 utterances from 28 speakers for training and 824 utterances from another 2 unseen speakers for testing. All utterances are resampled to 16KHz.

During the experiments, a Hamming window with 32ms window length and 8ms hop size (75\% overlap) is employed for both STFT and STDCT. Meanwhile, both the STFT and STDCT points are 512. The FME-Net contains five encoder blocks, three GRU layers, and five decoder blocks. The output channel of each convolutional layer in encoder is \{16,32,64,128,256\}, and correspondingly, the output channel of each deconvolutional layer in decoder is \{128,64,32,16,1\}. The kernel size and stride of all the (de)convolutional layers are set as (3,2) and (2,1), respectively. The hidden units of each GRU layer is \{128,64,32\}, and a fully-connected layer with 2304 units is after the last GRU layer. As for the DSR-Net, it also has five encoder blocks and five decoder blocks, and the setting of the output channels of the (de)convolutional layers is exactly the same as that of FME-Net. The convolutional stride is still (2,1), but the kernel size is changed to (5,2). DSR-Net includes three TFSM blocks, and the hidden GRU\&BiGRU units of each TFSM block is \{128,64,32\}. In order to ensure the real-time SE, all the (de)convolutional layers in our FDFNet is implemented causally by applying the asymmetric zero-padding. Furthermore, the FME-Net and DSR-Net have the same training strategy. In the training phase, RMSprop optimizer with the initial learning rate of 2e-4 is used. The learning rate will be halved if the model performance does not improve for five consecutive epochs. The batch size and total training epochs are 16 and 80.

\subsection{Ablation study}
\begin{table}[t]
  \centering
  \caption{Results of the ablation study}
  \setlength{\tabcolsep}{1.5mm}{
    \begin{tabular}{lccccc}
    \toprule[2pt]
          & TFSM & WB-PESQ & CSIG  & CBAK  & COVL \\
    \hline
    noisy & - & 1.97  & 3.35  & 2.44  & 2.63 \\
    \hline
    FME-Net & \XSolidBrush & 2.71 & 4.08 & 3.31 & 3.40 \\
    FME-Net$^{\dagger}$ & \CheckmarkBold & 2.81 & 4.09 & 3.37 & 3.45 \\
    DSR-Net$^{\star}$ & \XSolidBrush & 2.77 & 4.03 & 3.39 & 3.40 \\
    DSR-Net & \CheckmarkBold & 2.94 & 4.03 & 3.48 & 3.48 \\
    \hline
    FDFNet$^{\ast}$ & - & \textbf{3.05} & 4.21 & \textbf{3.55} & 3.64 \\
    FDFNet  & - & \textbf{3.05} & \textbf{4.23} & \textbf{3.55} & \textbf{3.65} \\
    \bottomrule[2pt]
    \end{tabular}}%
    \vspace{-0.4cm}
  \label{table1}%
\end{table}%

To demonstrate the effectiveness of our design, an ablation study is conducted as shown in Table\ \ref{table1}. We quantitatively evaluate the performance of model by a set of commonly used metrics, including wide-band perceptual evaluation of speech quality (WB-PESQ) \cite{941023}, and three MOS based metrics \cite{4389058} for signal distortion (CSIG), intrusiveness of background noise (CBAK), and overall audio quality (COVL).

We respectively test the performance of two single-stage models, i.e., FME-Net and DSR-Net. They are actually the CRN \cite{tan18_interspeech} and the DCTCRN \cite{li2021realtime} combined with our TFSM block. In addition, we also test the performance after adding our TFSM block to the FME-Net (marked as FME-Net$^{\dagger}$), as well as the performance after replacing the TFSM block in the DSR-Net with the GRU layer (marked as DSR-Net$^{\star}$). From the results, we can obtain the following observations: (1) Going from FME-Net (i.e., CRN) to DSR-Net$^{\star}$ (i.e., DCTCRN), the WB-PESQ and CBAK increase by 0.06 and 0.08, but the CSIG decreases by 0.05, which leads to the similar COVL. And same performance differences also exist between FME-Net$^{\dagger}$ (i.e., CRN+TFSM) and DSR-Net (i.e., DCTCRN+TFSM). This illustrates that although the STDCT-based model generally has superior performance, especially in noise suppression, it also causes more serious damage to the speech components. (2) The introduction of TFSM block improves the WB-PESQ, CSIG, CBAK, and COVL by 0.10, 0.01, 0.06, and 0.05 for FME-Net. As for DSR-Net, TFSM block also provides performance gain of 0.17 on WB-PESQ, 0.09 on CBAK, and 0.08 on COVL. This demonstrates the benefit of our TFSM block.

The two-stage model FDFNet is constructed by connecting FME-Net and DSR-Net together. In FDFNet, the input of DSR-Net includes the pre-enhanced spectrum of FME-Net, which has an obvious speech contour. Thus, this can guide the DSR-Net to better preserve the speech components. Meanwhile, the great noise reduction capability of the STDCT-based model ensures DSR-Net to effectively suppress the residual noise. In addition, the phase is also recovered by an implicit way in the STDCT spectrum domain. We find that the WB-PESQ, CSIG, CBAK, and COVL of FDFNet are respectively improved to 3.05, 4.23, 3.55, and 3.65. We have also tried to incorporate the TFSM block to FME-Net, which is marked as FDFNet$^{\ast}$. But this cannot further improve the model performance, which illustrates that in the two-stage pipeline, over-optimization for the first stage model may be unnecessary.

\subsection{Comparison with previous advanced systems}
\begin{table}[t]
  \centering
  \caption{Performance comparison with previous advanced systems under causal implementation.}
  \setlength{\tabcolsep}{0.45mm}{
    \begin{tabular}{lccccc}
    \toprule[2pt]
          & Param.(M) & WB-PESQ & CSIG  & CBAK  & COVL \\
    \hline
    noisy & - & 1.97  & 3.35  & 2.44  & 2.63 \\
    RNNoise\cite{8547084} & 0.06 & 2.29 & - & - & - \\
    ERNN\cite{9054597} & 0.79 & 2.54 & 3.74 & 2.65 & 3.13 \\
    DCCRN\cite{hu20g_interspeech} & 3.7 & 2.68 & 3.88 & 3.18 & 3.27 \\
    PerceptNet\cite{valin20_interspeech} & 8 & 2.73 & - & - & - \\
    DeepMMSE\cite{9066933} & - & 2.77 & 4.14 & 3.32 & 3.46 \\
    LFSFNet\cite{chen22c_interspeech} & 3.1 & 2.91 & - & - & - \\
    CTS-Net\cite{9431717} & 4.35 & 2.92 & 4.25 & 3.46 & 3.59 \\
    DEMUCS\cite{defossez20_interspeech} & 128 & 2.93 & 4.22 & 3.25 & 3.52 \\
    GaGNet\cite{LI2022108499} & 5.94 & 2.94 & \textbf{4.26} & 3.45 & 3.59 \\
    FDFNet  & 4.43 & \textbf{3.05} & 4.23 & \textbf{3.55} & \textbf{3.65} \\
    \bottomrule[2pt]
    \end{tabular}}%
    \vspace{-0.4cm}
  \label{table2}%
\end{table}%

We further compare our FDFNet with the previous advanced systems, and the results are presented in Table\ \ref{table2}. In these benchmarks, CTS-Net \cite{9431717} also adopts the two-stage pipeline, but its calculation process is only defined in the STFT spectrum domain. In CTS-Net, the second stage model adopts a dual-decoder network topology to capture a complex residual estimation, which is directly added to the first stage's output to obtain the final estimation. It can be observed that with similar parameter size, our FDFNet outperforms the CTS-Net on the WB-PESQ, CBAK, and COVL scores. As for the CSIG, our FDFNet is slightly lower than that of CTS-Net, which may be due to the fact that the speech damage caused by STDCT has not been fully compensated. Furthermore, with the same CRN structure and slightly fewer parameters, the DCTCRN (3.1M), i.e., the DSR-Net$^{\star}$ in Table\ \ref{table1}, outperforms DCCRN \cite{hu20g_interspeech} (3.7M). This further proves the advantage of STDCT over STFT. In addition, compared with other advanced systems, our FDFNet also illustrates superior performance. The enhanced audio clips can be found at \url{https://github.com/Zhangyuewei98/FDFNet.git}.


\section{Conclusions}
\label{sec:conclusions}
In this work, we improve the previous two-stage pipeline by the STFT-STDCT spectrum fusion network. The FDFNet first enhances the STFT magnitude spectrum, and then converts the pre-enhanced result into the STDCT spectrum domain for further residual noise suppression and implicit phase recovery. The experimental results demonstrate that our FDFNet outperforms the previous two-stage methods and other advanced systems. In the future, we will further optimize our scheme from the perspective of reducing speech distortion.

\section{Acknowledge}
This work was supported by the special funds of Shenzhen Science and Technology Innovation Commission under Grant No. CJGJZD20220517141400002.

\clearpage

\bibliographystyle{IEEEbib}
\bibliography{refs}

\end{document}